\documentclass[prb,twocolumn]{revtex4}%

\usepackage{graphicx}
\usepackage{amsmath}
\usepackage{ulem}
\usepackage{color}

\newcommand{\be}{\begin{equation}}
\newcommand{\ee}{\end{equation}}
\newcommand{\bea}{\begin{eqnarray*}}
\newcommand{\eea}{\end{eqnarray*}}
\newcommand{\bean}{\begin{eqnarray}}
\newcommand{\eean}{\end{eqnarray}}

\begin{document}

\draft
\title
{\bf High thermoelectric figure of merit of quantum dot array
quantum wires}

\author{David M T Kuo}
\address{Department of Electrical Engineering and Department of Physics, National Central
University, Chungli, 320 Taiwan}

\date{\today}

\begin{abstract}
How to design silicon-based quantum wires with figure of merit
($ZT$) larger than three is under hot pursuit due to the advantage
of low cost and the availability of matured fabrication technique.
Quantum wires consisting of finite three dimensional  quantum dot
(QD) arrays coupled to electrodes are proposed to realize high
efficient thermoelectric devices with optimized power factors. The
transmission coefficient of 3D QD arrays can exhibit 3D, 2D, 1D
and 0D topological distribution functions by tailoring the
interdot coupling strengths. Such topological effects on the
thermoelectric properties are revealed. The 1D topological
distribution function shows the maximum power factor and the best
$ZT$ value. We have demonstrated that 3D silicon QD array
nanowires with diameters below $20~nm$ and length $250~nm$ show
high potential to achieve $ZT\ge 3$ near room temperature.
\end{abstract}

\maketitle

\section{Introduction}
Thermoelectric power is one of the most important green energies,
which play a remarkable role for eternal development of earth.[1]
Unlike solar cells and wind-power generators, thermoelectric
devices have not only the functionality of power generators but
also coolers. Designing a thermoelectric material with a high
figure of merit ($ZT$) and optimized power output is under
pursuit.[2-5] Extensive studies have shown that it is difficult to
achieve a $ZT$ larger than one at room temperature in bulk
thermoelectric materials.[1] With the advances of semiconductor
technology, many experiments nowadays can realize $ZT$ larger than
one at room temperature in low-dimensional systems.[6,7] If a
material with $ZT \ge 3$ at room temperature it will tremendously
brighten the scenario of thermoelectric devices[1]. The
enhancement of power output call for a large number of electronic
states (bulk-like). However, a large $ZT$ value occurs in dilute
electronic states and phonon excitations (atom-like).[8]

It has been predicted theoretically  that  $ ZT \ge 3 $ can be
achieved in thin semiconductor nanowires with diameters smaller
than $3~nm$[9,10]. However, no experimental realization of such
impressive thermoelectric devices has been reported[1]. The
finding of $ZT=1$ in silicon nanowires at room temperature[11] has
inspired further studies of thermoelectric properties of
silicon-based nanowires because of the advantages of low cost and
the availability of matured fabrication technology in silicon
industry[11-14]. Whether $ZT \ge 3$ exists in silicon-based
nanowires at room temperature becomes an interesting topic.[14] It
is calculated that Si/Ge quantum dot suerlattice (QDSL) nanowire
with an optimized period around 5nm and cross-sectional area
around $3nm \times 3nm$ can lead to an reduction of phonon thermal
conductance by one order of magnitude in comparison with pristine
Si nanowires[15,16]. Therefore, it is desirable to study the power
output and $ZT$ of quantum wire consisting of finite 3D QD arrays
(QDAs) in ballistic transport procedure as depicted in Fig. 1.
Under tailoring interdot coupling strengths, the transmission
coefficient of finite size 3D QDAs can exhibit 3D, 2D, 1D and 0D
distribution functions. Their effects on the thermoelectric
properties of finite size QDA nanowires are examined. We found
that $ZT \ge 3 $ at room temperature may be achieved in a quantum
wire made of finite 3D Si/Ge QDAs with a diameter below $20~nm$
and a length $L_x=250~nm$.

\begin{figure}[h]
\centering
\includegraphics[trim=2.5cm 0cm 2.5cm 0cm,clip,angle=-90,scale=0.3]{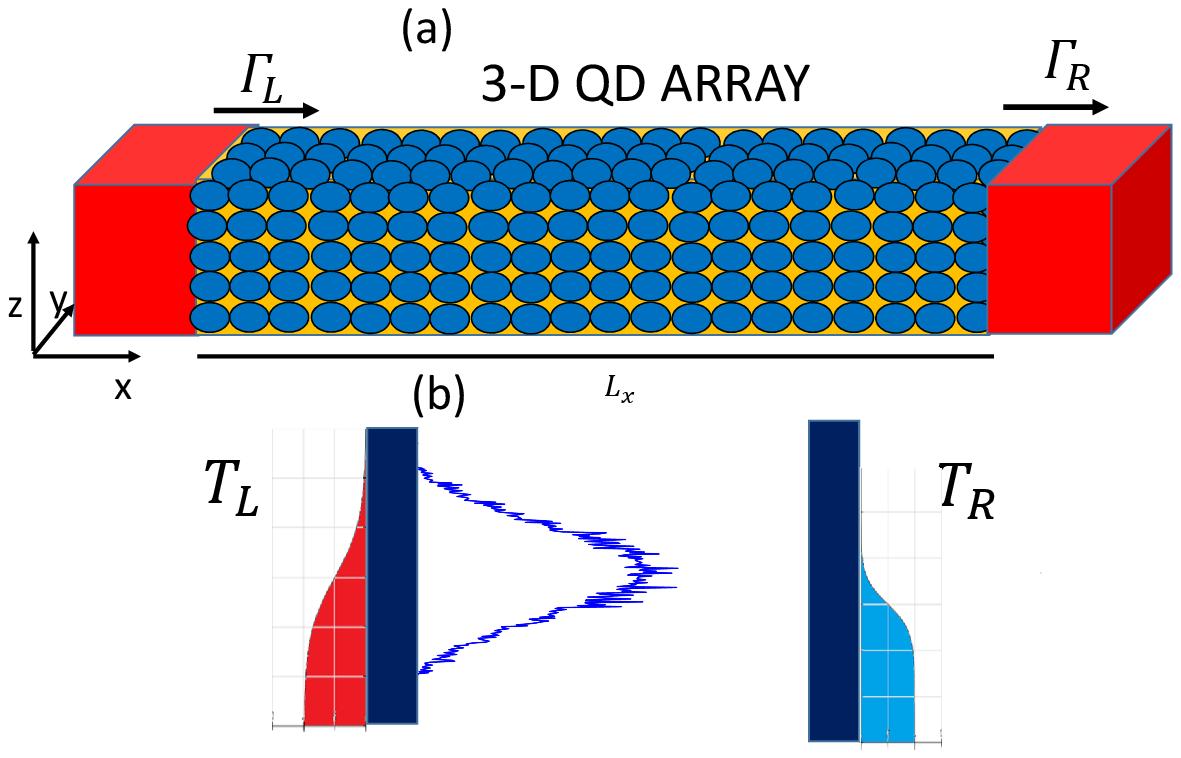}
\caption{(a) Schematic diagram of a quantum wire consisting of
finite 3D quantum dot (QD) array coupled to electrodes.
$\Gamma_{L}$ ($\Gamma_R$) denotes the tunneling rate of the
electrons between the left (right) electrode and the leftmost
(rightmost) QDs. (b) Electronic spectra of a 3D QD array coupled
to electrodes with different equilibrium temperatures ($T_L$ and
$T_R$). }
\end{figure}

\section{Formalism}
To model the thermoelectric properties of a finite size QDA
nanowire connected to the electrodes, the Hamiltonian of the
system shown in Fig. 1 is given by $H=H_0+H_{QD}$,[17] where
\begin{small}
\begin{eqnarray}
H_0& = &\sum_{k,\sigma} \epsilon_k
a^{\dagger}_{k,\sigma}a_{k,\sigma}+ \sum_{k,\sigma} \epsilon_k
b^{\dagger}_{k,\sigma}b_{k,\sigma}\\ \nonumber &+&\sum_{\ell}
\sum_{k,\sigma} V^L_{k,\ell}d^{\dagger}_{\ell,\sigma}a_{k,\sigma}
+\sum_{\ell}
\sum_{k,\sigma}V^R_{k,\ell}d^{\dagger}_{\ell,\sigma}b_{k,\sigma}+H.c.
\end{eqnarray}
\end{small}
The first two terms of Eq.~(1) describe the free electron gas in
the left and right electrodes. $a^{\dagger}_{k,\sigma}$
($b^{\dagger}_{k,\sigma}$) creates  an electron of momentum $k$
and spin $\sigma$ with energy $\epsilon_k$ in the left (right)
electrode. $V^L_{k,\ell}$ ($V^R_{k,\ell}$) describes the coupling
between the left (right) lead with its adjacent QD in the $\ell$th
site.

\begin{eqnarray}
H_{QD}&= &\sum_{\ell,\sigma} E_{0}
d^{\dagger}_{\ell,\sigma}d_{\ell,\sigma}+\sum_{\ell ,j,\sigma}
-t_{\ell,j} d^{\dagger}_{\ell, \sigma} d_{j,\sigma}+H.c,
\end{eqnarray}
where { $E_0$} is the site-independent energy level of QD. The
spin-independent $t_{\ell, j}$ describes the interdot coupling
strength, which is limited to the nearest neighboring sites.
$d^{\dagger}_{\ell,\sigma} (d_{\ell,\sigma})$ creates (destroys)
one electron in the $\ell$th site QD. If the wave functions of the
electrons in each QD are localized, the electron Coulomb
interactions are strong. Their effects on electron transport are
significant in the scenario of weak interdot coupling
strengths.[18] On the other hand, the wave functions of the
electrons are delocalized in the scenario of strong interdot
coupling strengths to form minibands; hence their weak electron
Coulomb interactions can be ignored. In eqs. (1) and (2), we take
into account one energy level for each QD due to nanoscale QDs
considered.

To study the transport properties of a QDA nanowire junction
connected to electrodes, it is convenient to use the
Keldysh-Green's function technique[17]. Electron and heat currents
leaving electrodes can be expressed as
\begin{eqnarray}
J &=&\frac{\sigma_s e}{h}\int {d\epsilon}~
T_{LR}(\epsilon)[f_L(\epsilon)-f_R(\epsilon)],
\end{eqnarray}
and
\begin{eqnarray}
& &Q_{e,L(R)}\\ &=&\frac{\pm \sigma_s}{h}\int {d\epsilon}~
T_{LR}(\epsilon)(\epsilon-\mu_{L(R)})[f_L(\epsilon)-f_R(\epsilon)]\nonumber
\end{eqnarray}
where
$f_{\alpha}(\epsilon)=1/\{\exp[(\epsilon-\mu_{\alpha})/k_BT_{\alpha}]+1\}$
denotes the Fermi distribution function for the $\alpha$-th
electrode, where $\mu_\alpha$  and $T_{\alpha}$ are the chemical
potential and the temperature of the $\alpha$ electrode. $e$, $h$,
and $k_B$ denote the electron charge, the Planck's constant, and
the Boltzmann constant, in that order. The factor $\sigma_s$
include electron spin and valley degeneracy of QDs.
$T_{LR}(\epsilon)$ denotes the transmission coefficient of a 3D
QDA connected to electrodes, which can be solved by the formula $
T_{LR}(\epsilon)=4Tr[\hat{\Gamma}_{L}\hat{G}^{r}_{QDA}(\epsilon)\hat{\Gamma}_{R}\hat{G}^{a}_{QDA}(\epsilon)]$,[19-21]
where the matrix of tunneling rates ($\hat{\Gamma}_L$ and
$\hat{\Gamma}_R$) and Green's functions
($\hat{G}^{r}_{QDA}(\epsilon)$ and $\hat{G}^{a}_{QDA}(\epsilon)$)
can be constructed by coding.[21]

The electrical conductance ($G_e$), Seebeck coefficient ($S$) and
electron thermal conductance ($\kappa_e$) can be evaluated by
using Eqs. (3) and (4) with a small applied bias $\Delta
V=(\mu_L-\mu_R)/e$ and cross-junction temperature difference
$\Delta T=T_L-T_R$. We obtain these thermoelectric coefficients
$G_e=e^2{\cal L}_{0}$, $S=-{\cal L}_{1}/(eT{\cal L}_{0})$ and
$\kappa_e=\frac{1}{T}({\cal L}_2-{\cal L}^2_1/{\cal L}_0)$. ${\cal
L}_n$ is given by
\begin{equation}
{\cal L}_n=\frac{\sigma_s}{h}\int d\epsilon~
T_{LR}(\epsilon)(\epsilon-\mu)^n\frac{-\partial
f(\epsilon)}{\partial \epsilon},
\end{equation}
where $f(\epsilon)=1/(exp^{(\epsilon-\mu)/k_BT}+1)$ is the Fermi
distribution function of electrodes at equilibrium temperature $T$
and chemical potential $\mu$. Factor $\sigma_s=12$ is for silicon
QDs.

\section{ Results and discussion}
To illustrate the thermoelectric properties of finite 3D QDAs, we
have calculated and shown in Fig. 2 transmission coefficient
$T_{LR}(\epsilon)$ as a function of $\epsilon$ for different
interdot coupling strength configurations at tunneling rate
$\Gamma_t=6\Gamma_0$ ($\Gamma_{L(R),\ell}(\epsilon)=2\pi\sum_{k}
|V^{L(R)}_{k,\ell}|^2 \delta(\epsilon-\epsilon_k)=\Gamma_t$) and
QDAs with $N_x=25$ and $N_y=N_z=N_r=5$. We employ $t_x$, $t_y$ and
$t_z$ to illustrate the interdot coupling strength $t_{\ell,j}$ in
the x, y and z directions, respectively. Diagram (a) considers
homogenous interdot coupling strengths $t_x=t_y=t_z=6\Gamma_0$ to
describe finite 3D QDAs. We turn off $t_z=0$ in diagram (b) to
illustrate multi-layer 2D structures. In diagram (c), $t_z=t_y=0$
describes multi 1D QDSLs. All energy scales of physical parameters
are in units of $\Gamma_0=1~meV$. Fig. 2 reveals the topological
effect on $T_{LR}(\epsilon)$ and the tunneling probability of the
electrons of the electrodes through the electronic states of QDAs.
Those electronic states are described by
$\epsilon=E_0-(2t_xcos(\frac{n_x\pi}{N_x+1})+
2t_ycos(\frac{n_y\pi}{N_y+1})+2t_zcos(\frac{n_z\pi}{N_z+1}))$,
where $n_x=1,2,..N_x$,$n_y=1,2,..N_y$ and $n_z=1,2,..N_z$. For the
situation of $t_x=t_y=t_z=6\Gamma_0$, a finite 3D QDA shows a 3D
topological distribution function in Fig. 2(a). When $t_z=0$, we
see a 2D topological distribution function in Fig. 2(b). We have a
typical 1D topological distribution function for $t_y=t_z=0$, as
seen in Fig. 2(c).[21] These electronic states have inhomogeneous
broadening due to the leftmost and rightmost QDs of 3D QDAs
connected to the electrodes. Electronic states near $E_0$ are more
broadening than those near band edges. For 1D topological
situation, lower band edge (LBE) and upper band edge (UBE) are,
respectively, $-12\Gamma_0$ and $12\Gamma_0$. The area between LBE
and UBE is so called the band regime (BR). For 2D and 3D
topological situations, their band widths (BWs) are $48$ and $72
\Gamma_0$, respectively. Physical parameters considered such as QD
energy level $E_0$, inter-dot coupling strength and tunneling rate
can be evaluated in the framework of effective mass model.[22]
According to eqs. (3) and (4), the maximum electron current and
heat current occur at $T_{LR}(\epsilon)$ with the maximum area at
a fixed BW. Our previous work has demonstrated that the condition
of $\Gamma_t=t_{x}$ has the optimization of $T_{LR}(\epsilon)$ of
QDAs.[21] Therefore, $\Gamma_t=t_x$ will be used through out this
article.

\begin{figure}[h]
\centering
\includegraphics[angle=-90,scale=0.3]{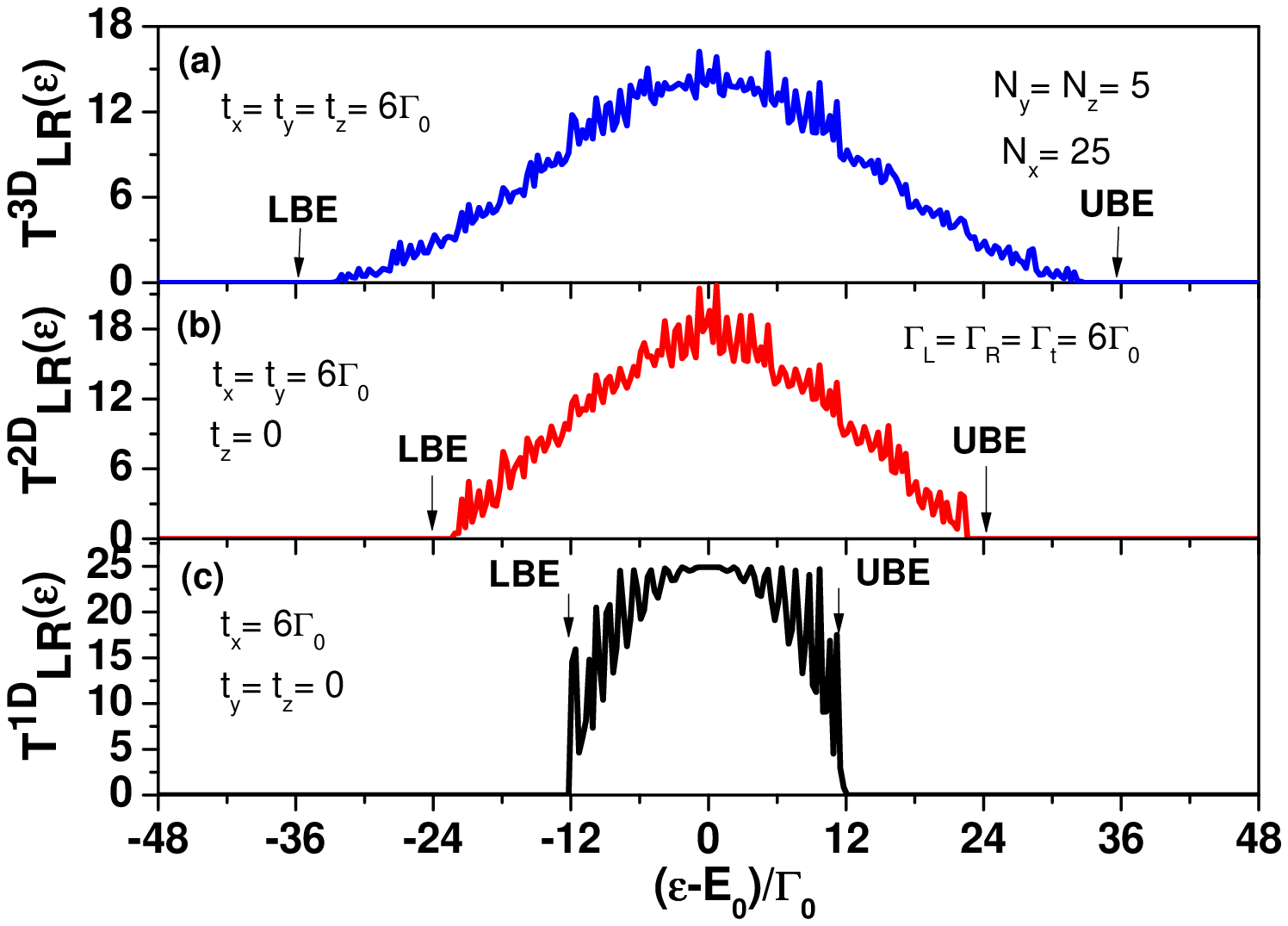}
\caption{Transmission coefficient $T_{LR}(\varepsilon)$ as
functions of $\varepsilon$ for different inter-dot coupling
strength configurations at $\Gamma_L=\Gamma_R=\Gamma_t=6\Gamma_0$
and $N_x=25$, $N_y=N_z=N_r=5$.  We set $E_0=0$ through out this
article.}
\end{figure}

Fig. 3 shows the calculated $G_e$, $S$ and power factor
($PF=S^2G_e$) at functions of $\mu$ for different topological
distribution functions shown in Fig.2 at two different
temperatures. Due to temperature effect, the highly oscillatory
electronic states shown in $T_{LR}(\varepsilon)$ are washed out in
the spectra of $G_e$. The maximum power factor in Fig. 3(c) is
given by the 1D topological case. The 1D maximum $PF$ values occur
at $|\mu|= 14 \Gamma_0$, where the 1D Seebeck coefficient is
significantly larger than that of 2D or 3D as seen in Fig 3(b).
This is because 1D transmission coefficient shows the steep change
with respect to $\epsilon$. In Fig. 3(f) the power factor is
almost independent on the topological situations as
$k_BT=25\Gamma_0$. Based on eq. (5), the range of integration
increases with increasing temperature. At $BW/(2k_BT)\le 1$ we
have $\int d\varepsilon T^{nD}_{LR}(\varepsilon)= C$ where $C$ is
a constant, this explains why the curves of $G_e$ and $S$ merge
together at high temperatures. In particular, $S\approx \mu/(eT)$
in Fig. 3(e). To further clarify the behaviors of Fig. 3 (d) and
(e), we assume transmission coefficient
$T_{LR}(\epsilon)=4\Gamma^2_t
N^2_{r}/((\epsilon-E_0)^2+(2\Gamma_t)^2)$ and $\Gamma_t/(2k_BT)\ll
1$ in eq. (5). $G_{e}=\frac{e^2}{h} \frac{N^2_{r}\pi \Gamma_t}
{2k_BT cosh^2(\mu/(2k_BT))}$ and $S=\mu/(eT)$ are obtained. The
results of Fig. 3(e) indicate that the thermoelectric behavior of
finite 3D QDAs with finite BWs is similar to that of a single QD
coupled to electrodes as $BW/(2k_BT)\le 1$. In Fig. 3, $G_e$ and
$S$ do not change at $k_BT=25\Gamma_0$ if we continuously increase
$N_x$ at a fixed $N_r=5$ (see red curves in Fig.6).

\begin{figure}[h]
\centering
\includegraphics[angle=-90,scale=0.3]{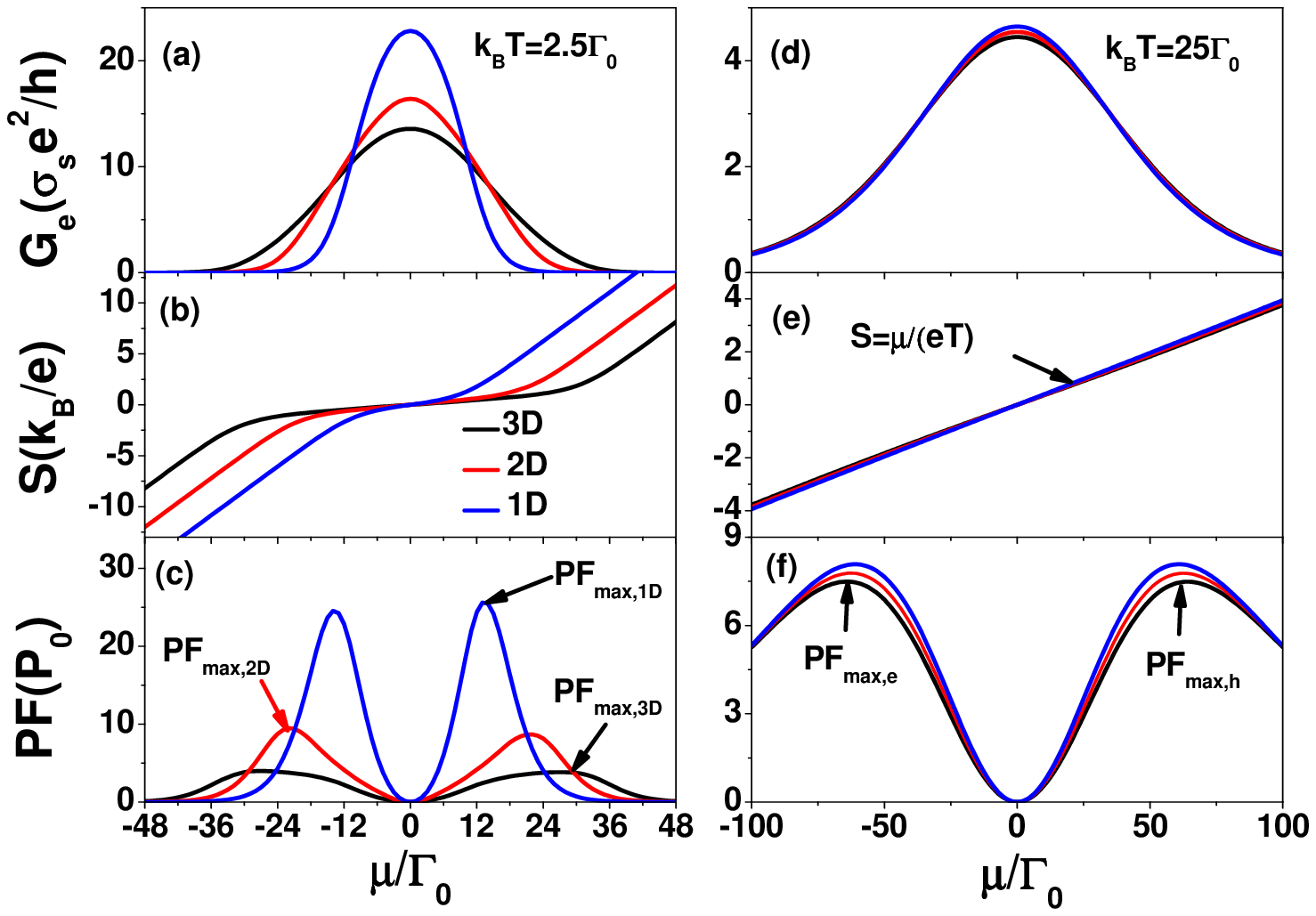}
\caption{(a) Electrical conductance $G_e$, (b) Seebeck coefficient
$S$ and (c) power factor $PF=S^2G_e$ as functions of $\mu$ for
three different topological spectra shown in Fig. 2 at
$k_BT=2.5\Gamma_0$. The curves of diagrams (d), (e) and (f)
consider the case of $k_BT=25\Gamma_0$. We have $P_0=
\sigma_sk^2_B/h^2$.}
\end{figure}

To further clarify the behavior of thermoelectric coefficients
with respect to temperature shown in Fig.3, we have calculated
$G_e$, $S$, $PF$ and $ZT$ as functions of $k_BT$ at
$\mu=-60\Gamma_0$ and $N_x=50$ in Fig. 4. In addition, we have
added extra curves to consider the 1D case with a smaller
$t_x=\Gamma_t=2\Gamma_0$ value, which is called atomic-limit
($0D$) to distinguish from the 1D case of
$t_x=\Gamma_t=6\Gamma_0$. Because only electronic states around
$\mu$ give the contributions of electron transport between the
electrodes at low temperatures, we see the vanishingly small
conductance at $k_BT\le 5\Gamma_0$ for $\mu=-60\Gamma_0$, which is
far away from the band center ($E_0$). Electron transport is
mainly contributed to thermionic assisted tunneling process (TATP)
as seen in Fig. 4(a). Due to the widest band width, $G_e$ shows
the maximum conductance in the 3D case. As for Seebeck
coefficients, it is expected that $S_{0D}=\mu/(eT)$ has a maximum
value. We note that $S_{0D}$ is determined only by $T$ as $\mu$ is
fixed. In particular, the dimensional effect on $S$ is disappear
at $k_BT\ge 25\Gamma_0$. This unique characteristic of $S$ may
exist a useful application of temperature sensors. When $k_BT >
10\Gamma_0$, the trend of $PF_{1D}> PF_{2D} > PF_{3D}$ is seen in
Fig. 4(c). The maximum power factor reaches $PF_{1D, max}=8P_0$ at
room temperature. This indicates that system provides the
electrical power output $7.82 nW/K$ at room temperature. Although
$S_{0D}$ has the largest value in the regime of $k_BT> 5
\Gamma_0$, its power factor is poor.

\begin{figure}[h]
\centering
\includegraphics[angle=-90,scale=0.3]{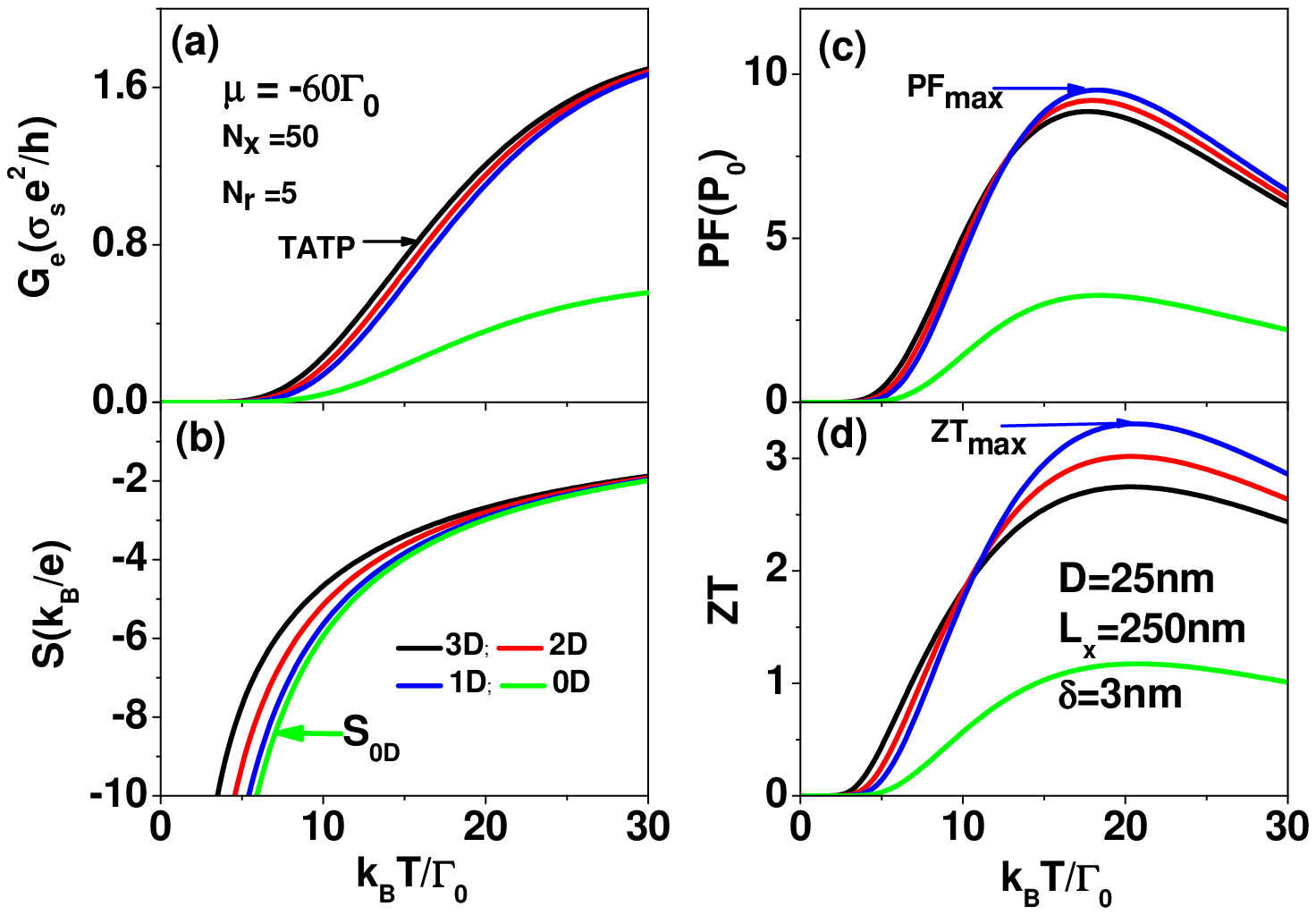}
\caption{(a) Electrical conductance, (b) Seeback coefficient, (c)
power factor $PF$ and (d) figure of merit $ZT$  as functions of
$k_BT$  at $\mu=-60\Gamma_0$, $N_r=5$ and $N_x=50$. To calculate
$\kappa_{ph}$, we have adopted diameter $D=25nm$, length
$L_x=250nm$ and surface roughness $\delta=3nm$.}
\end{figure}

The thermoelectric efficiency of materials is determined by the
figure of merit, $ZT=S^2G_eT/(\kappa_e+\kappa_{ph})$.
$\kappa_{ph}$ denotes the phonon thermal conductance of finite 3D
QDAs. It is difficult to fully block phonon heat currents because
QDs are embedded into matrix.[18] We should include $\kappa_{ph}$
in the calculation of $ZT$. We have adopted the following
empirical formula given by Ref.[23]

\begin{equation} \kappa_{ph}=\frac{F_s}{h} \int d\omega {\cal T}_{ph}(\omega) \frac {\hbar^3 \omega^2}{k_B T^2}\frac
{e^{\hbar\omega/k_BT}}{(e^{\hbar\omega/k_BT}-1)^2}, \label{phC}
\end{equation}
where $\omega$ and ${\cal T}_{ph}(\omega)$ are the phonon
frequency and throughput function, respectively. A dimensionless
factor $F_s$ is introduced to describe the reduction factor for
phonon transport due to scattering from QDs embedded in a quantum
wire. The value of $F_s=0.1$ is used, which is determined
according to Ref.[16], in which the phonon thermal conductance of
silicon/germanium  QD nanowires is calculated. As $F_s=1$, Eq. (6)
well explains the $\kappa_{ph}$ of silicon quantum wires with
diameters $D\ge 20~nm$.[23,24]. In Fig. 4(d) we have considered a
quantum wire with diameter $D=25~nm$ and length $L_x=250~nm$ since
the pair length (one QD plus spacer layer) adopted is
$L_s=5nm$.[16] The maximum $ZT$ is given by $(ZT)_{1D}=3.31$ at
$k_BT=20\Gamma_0$. This is a remarkable result, since quantum
wires with $ZT \ge 3$ have been theoretically reported only for $D
< 3~nm$.[1,9] So far, we are limited to the situation of
$N_z=N_y=N_r=5$. Although the power factor can be enhanced with
increasing $N_r$, $\kappa_{ph}$ is also highly enhanced with
increasing $D$.[23] We show the calculated $\kappa_{ph}$ as
functions of temperature for different diameters at $L_x=250~nm$
in Fig. 5. The line marked with triangles denotes the $\kappa_e$
corresponding to the curves of 3D case in Fig. 4. $\kappa_{ph}$
fully dominates heat current at room temperature
($k_BT=25\Gamma_0$). It is worthy noting that $\kappa_e >
\kappa_{ph}$ is observed in the situation of extremely low
temperatures $k_BT < 0.1\Gamma_0$. Many literatures have reported
the Carnot engine behavior of a single QD when $\kappa_e \gg
\kappa_{ph}$.[25-27].

\begin{figure}[h]
\centering
\includegraphics[angle=-90,scale=0.3]{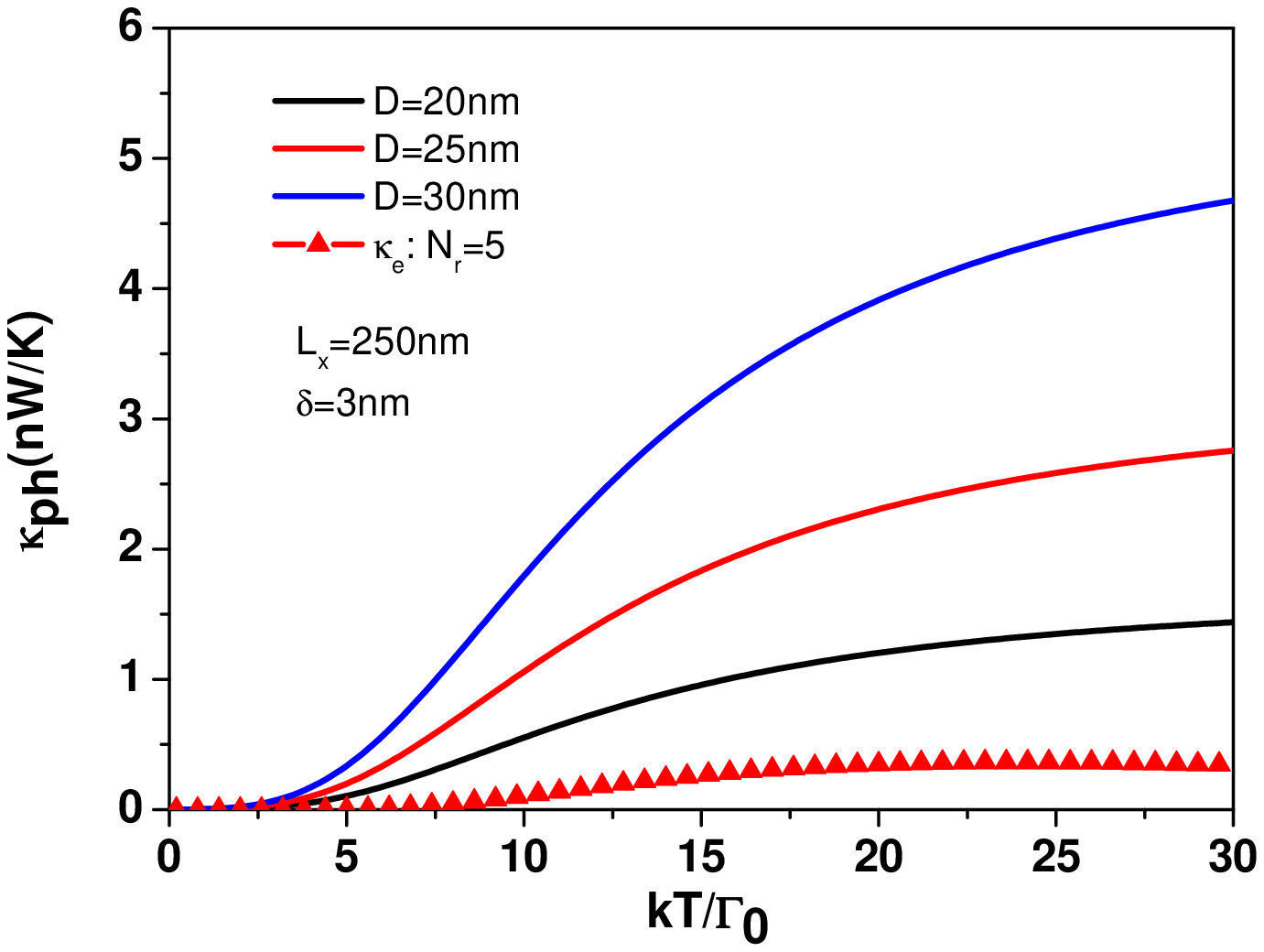}
\caption{Phonon thermal conductance $\kappa_{ph}$ as functions of
temperature for different diameters at length $L_x=250~nm$ and
surface roughness $\delta=3~nm$.}
\end{figure}

It is important to investigate the effect of quantum wire diameter
on $ZT$ because $\kappa_{ph}$ is a function of $D$. We show the
calculated $G_e$, $S$, $PF$ and $ZT$ as functions of $\mu$ for
different $N_r$ values at $k_BT=25\Gamma_0$, $N_x=50$ and
$t_x=t_y=t_z=\Gamma_t=6\Gamma_0$ in Fig. 6. From experimental
point of view, it is relatively easy to fabricate the case of
$t_x=t_y=t_z$.[28,29] For comparison, $0D$ case is also included
in Fig. 6. $G_e$ is highly enhanced with increasing $N_r$, whereas
the enhancement of $G_e$ does not suppress $S$. As a consequence,
$PF$ is enhanced with increasing $N_r$. This crucial
characteristic is meaningful to increase the electrical power
output. Quantum wire diameters are, respectively, $D=20, 25 $ and
$30~nm$ for $N_r=4,5$ and $6$. In Fig. 6(d) the maximum $ZT=3.31$
is given by $D=20nm$ and $\mu=-70\Gamma_0$. Note that the curve of
a QDA nanowire exhibiting a 0D topological electron distribution
function exists poor $PF$ and $ZT$ values when phonon throughput
function maintains a 3D distribution function characteristic. In
Ref.[2], $T^{0D}_{LR}(\epsilon)$ shows the Carnot efficiency
because they assumed that the condition of $\kappa_{e}\gg
\kappa_{ph}$ can be realized experimentally.
\begin{figure}[h]
\centering
\includegraphics[angle=-90,scale=0.3]{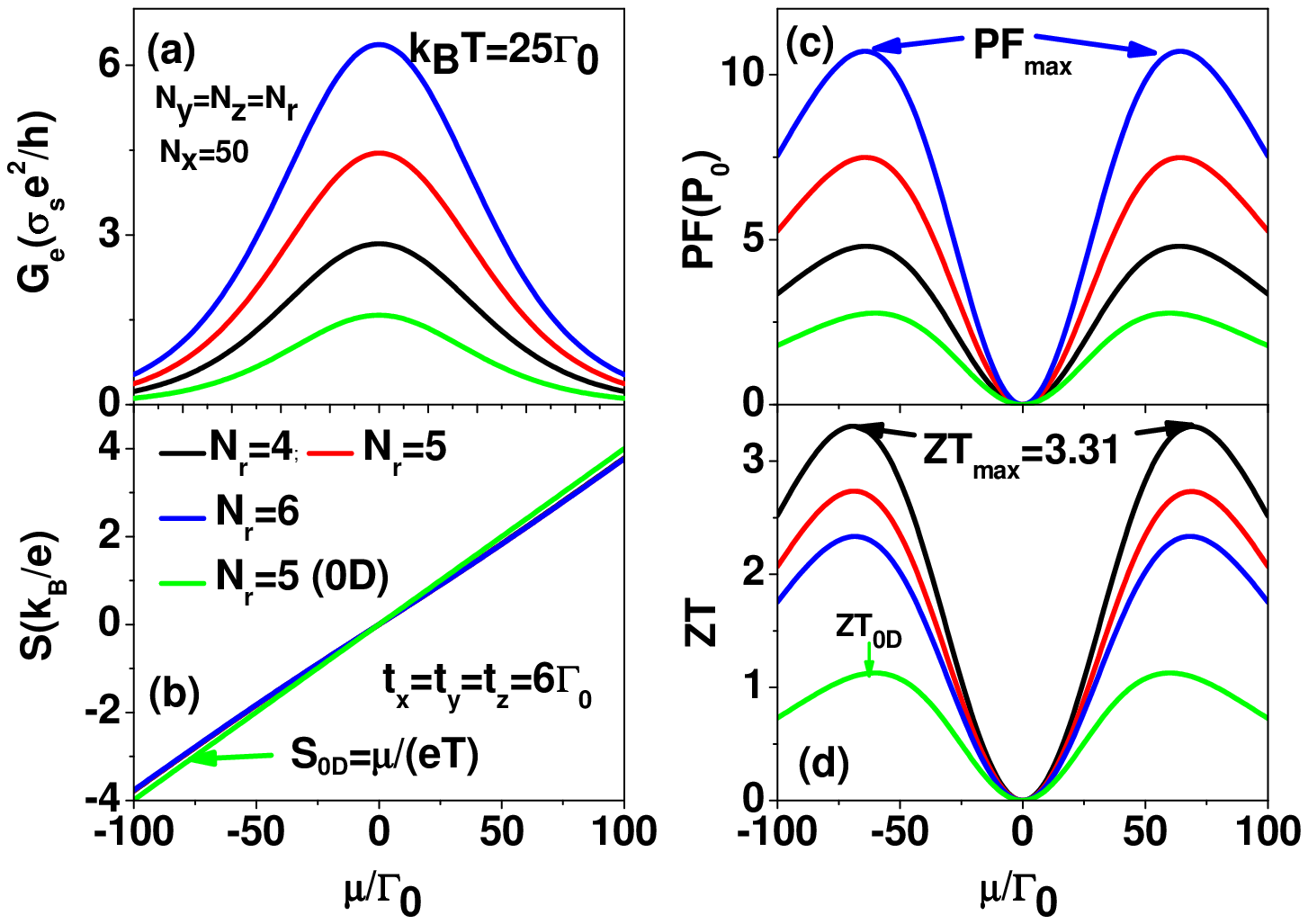}
\caption{(a) Electrical conductance, (b) Seeback coefficient, (c)
power factor $PF$ and (d) figure of merit $ZT$  as functions of
$\mu$ for various $N_r$ values at $t_x=t_y=t_z=6\Gamma_0$. In the
calculation of $\kappa_{ph}$ , we have adopted $D=20,25$ and
$30~nm$ for $N_r=4,5$ and $6$, respectively. The curve of $0D$
corresponds to that of Fig. 4. Other physical parameters are the
same as those of Fig.4.}
\end{figure}

\section{Conclusion}
We have theoretically investigated the thermoelectric properties
of QDA quantum wires in the ballistic transport regime. According
to Ref.[30], the electron mean free path $\lambda_e$ reaches $300
nm$ at room temperature in Si nanowires. Therefore, it is adequate
to consider electron transport in the ballistic regime at $L_x \le
250~nm$. The interdot coupling strengths $t_{\ell,j}$ are
determined by the barrier width and height, which can be designed
on the demand. For example, if Si QDs have $SiO_2$ barrier widths,
$t_{\ell,j}$ are vanishingly small. Tailoring $t_{\ell,j}$,
$T_{LR}(\epsilon)$ exhibits the 3D, 2D, 1D and 0D topological
distribution functions. We have revealed such topological effects
on the power factor and figure of merit of quantum wires. Although
$T^{1D}_{LR}(\epsilon)$ provides the maximum power factor and the
largest $ZT$, $T^{3D}_{LR}(\epsilon)$ still shows a $ZT > 3$ for a
finite size silicon QDA nanowire with a diameter $D=20~nm$ and a
length $L_x=250~nm$ in the TATP.



{\bf Acknowledgments}\\
This work was supported under Contract No. MOST
110-2119-M-008-006-MBK
\mbox{}\\
E-mail address: mtkuo@ee.ncu.edu.tw\\

\setcounter{section}{0}

\renewcommand{\theequation}{\mbox{A.\arabic{equation}}} 
\setcounter{equation}{0} 

\mbox{}\\



\end{document}